\documentclass{ws-ijmpe}

\begin{document}

\markboth{LIE-WEN CHEN, CHE MING KO, and BAO-AN LI}{Constraining
the Skyrme effective interactions and the neutron skin thickness
of nuclei}

\catchline{}{}{}{}{}

\title{CONSTRAINING THE SKYRME EFFECTIVE INTERACTIONS
AND THE NEUTRON SKIN THICKNESS OF NUCLEI USING ISOSPIN DIFFUSION
DATA FROM HEAVY ION COLLISIONS}

\author{LIE-WEN CHEN}

\address{Institute of Theoretical Physics, Shanghai Jiao Tong University, \\
Shanghai 200240, China\\
lwchen@sjtu.edu.cn}

\author{CHE MING KO}

\address{Cyclotron Institute and Physics Department, Texas A\&M University, \\
College Station, Texas 77843-3366, USA\\
ko@comp.tamu.edu}

\author{BAO-AN LI}

\address{Department of Physics, Texas A\&M University-Commerce, \\
Commerce, Texas 75429, and Department of Chemistry and Physics, \\
P.O. Box 419, Arkansas State University, State University, Arkansas 72467-0419, USA \\
Bao-An\_Li@TAMU-Commerce.edu}

\maketitle

\begin{history}
\received{(received date)}
\revised{(revised date)}
\end{history}

\begin{abstract}
Recent analysis of the isospin diffusion data from heavy-ion
collisions based on an isospin- and momentum-dependent transport
model with in-medium nucleon-nucleon cross sections has led to the
extraction of a value of $L=88\pm 25$ MeV for the slope of the
nuclear symmetry energy at saturation density. This imposes
stringent constraints on both the parameters in the Skyrme
effective interactions and the neutron skin thickness of heavy
nuclei. Among the 21 sets of Skyrme interactions commonly used in
nuclear structure studies, the 4 sets SIV, SV, G$_\sigma$, and
R$_\sigma$ are found to give $L$ values that are consistent with
the extracted one. Further study on the correlations between the
thickness of the neutron skin in finite nuclei and the nuclear
matter symmetry energy in the Skyrme Hartree-Fock approach leads
to predicted thickness of the neutron skin of $0.22\pm 0.04$ fm
for $^{208}$Pb, $0.29\pm 0.04$ fm for $^{132}$Sn, and $0.22\pm
0.04$ fm for $^{124}$Sn.
\end{abstract}

\section{Introduction}

The study of the equation of state (EOS) of isospin asymmetric
nuclear matter, especially the nuclear symmetry energy, is
currently an active field of research in nuclear physics \cite%
{ireview98,ibook,bom,diep03,pawel02,lat04,baran05,steiner05}.
Although the nuclear symmetry energy at normal nuclear matter
density is known to be around $30$ \textrm{MeV} from the empirical
liquid-drop mass formula \cite{myers,pomorski}, its values at other
densities, especially at supranormal densities, are poorly known
\cite{ireview98,ibook}. Advances in radioactive nuclear beam
facilities provide, however, the possibility to pin down the density
dependence of the nuclear symmetry energy in heavy ion collisions
induced by these nuclei \cite{ireview98,ibook,baran05,li97,muller95,fra1,fra2,xu00,tan01,bar02,betty,lis,li00,npa01,li02,chen03,ono03,liu03,chen04,li04a,shi03,li04prc,rizzo04,liyong05,lizx05a,lizx05b,tian05,li06plb}%
. Indeed, significant progress has recently been made in
extracting the information on the density dependence of nuclear
symmetry energy from the isospin diffusion data in heavy-ion
collisions at NSCL/MSU \cite{tsang04,chen05,li05}. Using an
isospin- and momentum-dependent IBUU04 transport model with
in-medium nucleon-nucleon (NN) cross sections, the isospin
diffusion data were found to be consistent with a relatively soft
nuclear symmetry energy at subnormal density \cite{li05}.

Information on the density dependence of the nuclear symmetry
energy can also be obtained from the thickness of the neutron skin
in heavy nuclei as the latter is strongly correlated with the
slope $L$ of the nuclear matter symmetry energy at saturation
density \cite{brown00,hor01,typel01,furn02,kara02,diep03}. Because
of the large uncertainties in measured neutron skin thickness of
heavy nuclei, this has not been possible. Instead, studies have
been carried out to use the extracted nuclear symmetry energy from
the isospin diffusion data to constrain the neutron skin thickness
of heavy nuclei \cite{steiner05b,li05}. Using the Hartree-Fock
approximation with parameters fitted to the phenomenological EOS
that was used in the IBUU04 transport model to describe the
isospin diffusion data from NSCL/MSU, it was found that a neutron
skin thickness of less than $0.15$ fm \cite{steiner05b,li05} for
$^{208}$Pb was incompatible with the isospin diffusion data.

In the present talk, we report our recent work on constraining the
Skyrme effective interactions and the neutron skin thickness of
heavy nuclei using the isospin diffusion data from heavy ion
collisions \cite{chen05nskin}. Using the value of $L$ obtained
from the extracted density dependence of the nuclear symmetry
energy from the isospin diffusion data, we have been able to limit
the allowed parameter sets for the Skyrme interaction. Also,
studying the correlation between the density dependence of the
nuclear symmetry energy and the thickness of the neutron skin in a
number of nuclei within the framework of the Skyrme Hartree-Fock
approach further allows us to obtain stringent
constraints on the neutron skin thickness of the nuclei $^{208}$Pb, $^{132}$%
Sn, and $^{124}$Sn.

\section{Nuclear symmetry energy and the Skyrme interaction}

The nuclear symmetry energy $E_{\text{sym}}(\rho )$ at nuclear
density $\rho$ can be expanded around the nuclear matter
saturation density $\rho _{0}$ as
\begin{equation}
E_{\text{sym}}(\rho )=E_{\text{sym}}(\rho _{0})+\frac{L}{3}\left(
\frac{\rho -\rho _{0}}{\rho _{0}}\right)
+\frac{K_{\text{sym}}}{18}\left( \frac{\rho -\rho _{0}}{\rho
_{0}}\right) ^{2},  \label{EsymLK}
\end{equation}%
where $L$ and $K_{\text{sym}}$ are the slope and curvature of the
nuclear symmetry energy at $\rho _{0}$, i.e.,
\begin{equation}
L=3\rho _{0}\frac{\partial E_{\text{sym}}(\rho )}{\partial \rho
}|_{\rho
=\rho _{0}},\text{ }K_{\text{sym}}=9\rho _{0}^{2}\frac{\partial ^{2}E_{\text{%
sym}}(\rho )}{\partial ^{2}\rho }|_{\rho =\rho _{0}}.
\label{LKsym}
\end{equation}%
The $L$ and $K_{\text{sym}}$ characterize the density dependence
of the nuclear symmetry energy around normal nuclear matter
density, and thus provide important information on the properties
of nuclear symmetry energy at both high and low densities.

In the standard Skyrme Hartree-Fock approach
\cite{brack85,fried86,brown98,clw99,stone03,brown00}, the
interaction is taken to be zero-range and density- and
momentum-dependent with its parameters fitted to the binding
energies and charge radii of a large number of nuclei in the
periodic table. For infinite nuclear matter, the nuclear symmetry
energy from the Skyrme interaction can be expressed as
\cite{clw99,stone03}
\begin{eqnarray}
E_{\text{sym}}(\rho ) &=&\frac{1}{3}\frac{\hbar ^{2}}{2m}\left(
\frac{3\pi
^{2}}{2}\right) ^{2/3}\rho ^{2/3}-\frac{1}{8}t_{0}(2x_{0}+1)\rho -\frac{1}{48%
}t_{3}(2x_{3}+1)\rho ^{\sigma +1}  \notag \\
&&+\frac{1}{24}\left( \frac{3\pi ^{2}}{2}\right)
^{2/3}[-3t_{1}x_{1}+\left( 4+5x_{2}\right) t_{2}]\rho ^{5/3},
\label{EsymSky}
\end{eqnarray}%
where the $\sigma $, $t_{0}-t_{3}$, and $x_{0}-x_{3}$ are Skyrme
interaction parameters.

\begin{figure}[th]
\centerline{\epsfig{file=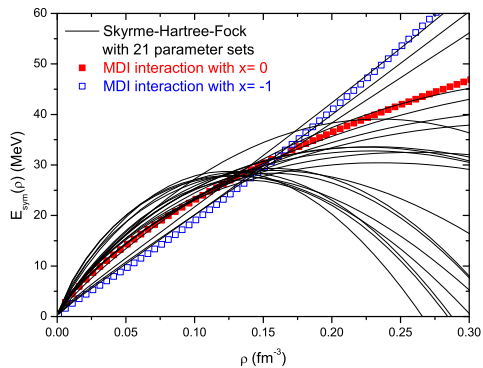,width=6.5cm}}
\vspace*{8pt} \caption{{\protect\small (Color online) Density
dependence of the nuclear symmetry energy
}$E_{\text{sym}}(\protect\rho )${\protect\small \ for 21 sets of
Skyrme interaction parameters (solid lines) as well as the MDI
interaction with }$x=-1${\protect\small \ (open squares) and
}$0${\protect\small \ (solid squares).}} \label{SymDen}
\end{figure}

Fig. \ref{SymDen} displays the density dependence of
$E_{\text{sym}}(\rho )$ for $21$ sets of Skyrme interaction parameters, i.e., \textrm{SKM}, \textrm{%
SKM}$^{\ast }$, $\mathrm{RATP}$, \textrm{SI}, \textrm{SII},
$\mathrm{SIII}$, \textrm{SIV}, \textrm{SV}, \textrm{SVI},
\textrm{E}, \textrm{E}$_{\sigma }$, \textrm{G}$_{\sigma }$, $\mathrm{R}_{\sigma }$, $\mathrm{Z}$, $\mathrm{Z}%
_{\sigma }$, $\mathrm{Z}_{\sigma }^{\ast }$, $\mathrm{T}$, $\mathrm{T3}$, $%
\mathrm{SkX}$, $\mathrm{SkXce}$, and $\mathrm{SkXm}$. Values of
the parameters in these Skyrme interactions can be found in Refs. \cite%
{brack85,fried86,brown98}. For comparison, we also show in Fig.
\ref{SymDen} results from the phenomenological parametrization of
the momentum-dependent nuclear mean-field potential based on the Gogny effective interaction \cite%
{das03}, i.e., the MDI interactions with $x=-1$ (open squares) and
$0$ (solid squares), where different $x$ values correspond to
different density dependence of the nuclear symmetry energy but
keep other properties of the nuclear EOS the same \cite{chen05}.
By comparing the isospin diffusion data from NSCL/MSU with results
from the IBUU04 transport model using in-medium NN cross sections,
these interactions have been shown to give,
respectively, the upper and lower bounds for the stiffness of the nuclear symmetry energy \cite%
{li05}. It is seen from Fig. \ref{SymDen} that the density
dependence of the symmetry energy varies drastically among
different interactions. Although the values of
$E_{\text{sym}}(\rho _{0})$ are all in the small range of
$26$-$35$ MeV, the values of $L$ and $K_{\text{sym}}$ are in a
much larger range of $-50$-$100$ MeV and $-700$-$50$ MeV,
respectively.

\section{Constraining symmetry energy from isospin
diffusion data in heavy ion collisions}

Nuclear symmetry energy is known to affect the diffusion of
isospin in intermediate-energy heavy ion collisions.
Experimentally, the degree of isospin diffusion between the
projectile nucleus $A$ and the target nucleus $B$ can be studied via the quantity $%
R_{i} $ \cite{rami,tsang04},
\begin{equation}
R_{i}=\frac{2X^{A+B}-X^{A+A}-X^{B+B}}{X^{A+A}-X^{B+B}}, \label{Ri}
\end{equation}%
where $X$ is any isospin-sensitive observable. By construction, the
value of $R_{i}$ is $1~(-1)$ for symmetric $A+A~(B+B)$ reaction. If
isospin equilibrium is reached during the collision as a result of
isospin diffusion, the value of $R_{i}$ is about zero. In the
NSCL/MSU experiments with $A=$ $^{124}$Sn and $B=$ $^{112}$Sn at a
beam energy of $50$ MeV/nucleon and an impact parameter about $6$
fm, the isospin asymmetry of the projectile-like residue was used as the isospin tracer $X$ \cite{tsang04}%
. Using an isospin- and momentum-dependent IBUU04 transport model
with free-space experimental NN cross sections, the dependence of
$R_{i}$ on the nuclear symmetry energy was studied from the
average isospin asymmetry of the projectile-like residue that was
calculated from nucleons with local densities higher than $\rho
_{0}/20$ and velocities larger than $1/2$ the beam velocity in the
center-of-mass frame \cite{chen05}. Comparing theoretical results
with experimental data has allowed us to extract a nuclear
symmetry energy of $E_{\text{sym}}(\rho )\approx 31.6(\rho /\rho
_{0})^{1.05}$. Including also medium-dependent NN cross sections,
which are important for isospin-dependent observables
\cite{li05,li05a}, the isospin diffusion data leads to an even softer nuclear symmetry energy of $E_{\text{%
sym}}(\rho )\approx 31.6(\rho /\rho _{0})^{\gamma }$ with $\gamma
\approx 0.7 $ \cite{li05}.

\begin{figure}[th]
\centerline{\epsfig{file=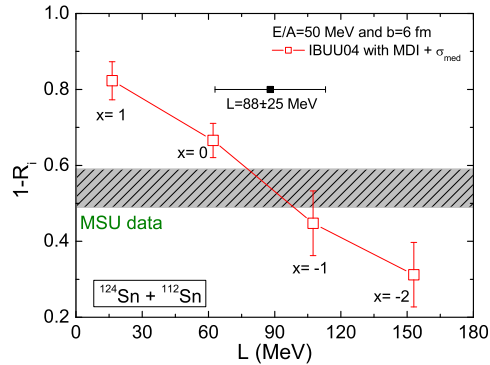,width=6.5cm}}
\vspace*{8pt}
\caption{{\protect\small (Color online) Degree of the isospin diffusion }$%
1-R_{i}${\protect\small \ as a function of }$L${\protect\small \
using the
MDI interaction with }$x=-2${\protect\small , }$-1${\protect\small , }$0$%
{\protect\small , and }$1${\protect\small . The shaded band
indicates data from NSCL/MSU \protect\cite{tsang04}. The solid
square with error bar represents }$L=88\pm 25${\protect\small \
MeV.}} \label{RiL}
\end{figure}

In Fig. \ref{RiL}, we show the degree of the isospin diffusion
$1-R_{i}$ as a function of $L$ obtained from the IBUU04 transport
model with in-medium NN cross sections and the mean-field
potential based on the MDI interactions. The shaded band in Fig.
\ref{RiL} indicates the data from NSCL/MSU \cite{tsang04}. It is
seen that the strength of isospin diffusion $1-R_{i}$ decreases
monotonically with decreasing value of $x$ or increasing value of
$L$. This is expected as the parameter $L$ reflects the difference
in the pressures on neutrons and protons. From comparison of the
theoretical results with the data, we can clearly exclude the MDI
interaction with $x=1$ and $x=-2$ as they give either too large or
too small a value for $1-R_{i}$ compared to that of data. The
range of $x$ or $L$ values that give values of $1-R_{i}$ falling
within the band of experimental values could in principle be
determined in our model by detailed calculations. Instead, we
determine this schematically by using the results from the four
$x$ values. For the centroid value of $L$, it is obtained from the
interception of the line connecting the theoretical results at
$x=-1$ and $0$ with the central value
of $1-R_{i}$ data in Fig. \ref{RiL}, i.e., $L=88$ MeV. The upper limit of $%
L=113$ MeV is estimated from the interception of the line connecting
the upper error bars of the theoretical results at $x=-1$ and $-2$
with the
lower limit of the data band of $1-R_{i}$. Similarly, the lower limit of $%
L=65$ MeV is estimated from the interception of the line connecting
the lower error bars of the theoretical results at $x=0$ and $-1$
with the upper
limit of the data band of $1-R_{i}$. This leads to an extracted value of $%
L=88\pm 25$ MeV as shown by the solid square with error bar in Fig. \ref{RiL}%
.

The extracted value of $L=88\pm 25$ MeV gives a rather stringent
constraint on the density dependence of the nuclear symmetry
energy and thus puts strong limitations on the nuclear effective
interactions as well. For the Skyrme effective interactions shown
in Fig. \ref{SymDen}, for instance, all of those lie beyond $x=0$
and $x=-1$ in the sub-saturation region are not consistent with
the extracted value of $L$. Actually, we note that
only $4$ sets of Skyrme interactions, i.e., $\mathrm{SIV}$, $\mathrm{SV}$, $\mathrm{G}%
_{\sigma }$, and $\mathrm{R}_{\sigma }$, in the $21$ sets of Skyrme
interactions considered here have nuclear symmetry energies that are
consistent with the extracted $L$ value.

\section{Correlations between neutron skin thickness of finite nuclei
and the nuclear symmetry energy at saturation density}

\begin{figure}[th]
\centerline{\epsfig{file=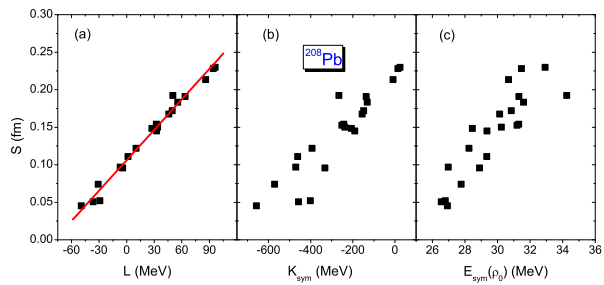,width=8cm}}
\vspace*{8pt}
\caption{{\protect\small (Color online) {Neutron skin thickness $S$ of $%
^{208}$Pb as a function of (a) $L$, (b) $K_{\text{sym}}$, and (c) $E_{\text{%
sym}}(\protect\rho _{0})$ for 21 sets of Skyrme interaction
parameters. }The line in panel (a) represents a linear fit.}}
\label{SPb208}
\end{figure}

Also affected by the density dependence of nuclear symmetry energy
is the neutron skin thickness $S$ of a nucleus, which is defined
as the difference between the root-mean-square radii
$\sqrt{\left\langle r_{n}\right\rangle }$ of neutrons and
$\sqrt{\left\langle r_{p}\right\rangle }$ of protons, i.e.,
\begin{equation}
S=\sqrt{\left\langle r_{n}^{2}\right\rangle }-\sqrt{\left\langle
r_{p}^{2}\right\rangle }.  \label{S}
\end{equation}%
In particular, $S$ is sensitive to the slope parameter $L$ of the
nuclear symmetry energy at normal nuclear matter density
\cite{brown00,hor01,typel01,furn02,kara02,diep03}. Using above
$21$ sets of Skyrme interaction parameters, we have evaluated the
neutron skin thickness of several nuclei. In Figs.
\ref{SPb208}(a), (b) and (c), we show, respectively, the
correlations between the neutron skin thickness of $^{208}$Pb with
$L$, $K_{\text{sym}}$, and $E_{\text{sym}}(\rho _{0})$. It is seen
from Fig. \ref{SPb208}(a) that there exists an approximate linear
correlation between $S$ and $L$. The correlations of $S$ with
$K_{\text{sym}}$ and $E_{\text{sym}}(\rho _{0})$ are less strong
and even exhibit some irregular behavior. The solid line in Fig.
\ref{SPb208}(a) is a linear fit to the correlation between $S$ and
$L$ and is given by the following expression:
\begin{equation}
S(^{\text{208}}\text{Pb)}=(0.1066\pm 0.0019)+(0.00133\pm
3.76\times 10^{-5})\times L,  \label{SLPb208a}
\end{equation}%
or
\begin{equation}
L=(-78.5\pm 3.2)+(740.4\pm 20.9)\times S(^{\text{208}}\text{Pb)},
\label{SLPb208b}
\end{equation}%
where the units of $L$ and $S$ are \textrm{MeV} and \textrm{fm},
respectively. Therefore, if the value for either
$S(^{\text{208}}$Pb) or $L$ is known, the value for the other can be
determined.

\begin{figure}[th]
\centerline{\epsfig{file=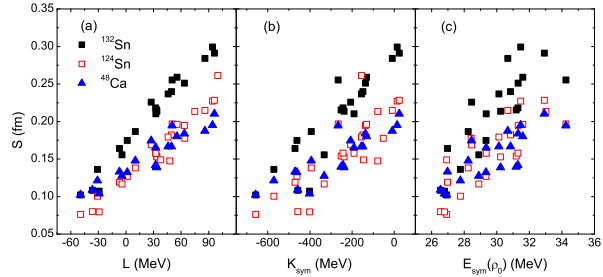,width=8cm}}
\vspace*{8pt} \caption{{\protect\small (Color online) {Same as
Fig. \ref{SPb208} but for nuclei $^{132}$Sn (Solid squares),
$^{124}$Sn (Open squares) and $^{48}$Ca (Triangles).}}}
\label{SSnCa}
\end{figure}

It is of interest to see if there are also correlations between
the neutron skin thickness of other neutron-rich nuclei and the
nuclear symmetry energy. Fig. \ref{SSnCa} shows the same
correlations as in Fig. \ref{SPb208} but for the neutron-rich
nuclei $^{132}$Sn, $^{124}$Sn, and $^{48}$Ca. For the heavier
$^{132}$Sn and $^{124}$Sn, we obtain a similar conclusion as for
$^{208}$Pb, namely, $S$ exhibits an approximate linear correlation
with $L$ but weaker correlations with $K_{\text{sym}}$ and
$E_{\text{sym}}(\rho _{0})$. For the lighter $^{48}$Ca, on the
other hand, all the correlations become weaker than those of
heavier nuclei. The neutron skin thickness of heavy nuclei is thus
better correlated with the density dependence of the nuclear
symmetry energy. As in Eq. (\ref{SLPb208a}) and (\ref{SLPb208b}),
a
linear fit to the correlation between $S$ and $L$ can also be obtained for $^{132}$%
Sn and $^{124}$Sn, and the corresponding expressions are
\begin{equation}
S(^{\text{132}}\text{Sn)}=(0.1694\pm 0.0025)+(0.0014\pm 5.12\times
10^{-5})\times L,  \label{SLSn132a}
\end{equation}%
\begin{equation}
L=(-117.1\pm 5.4)+(695.1\pm 25.3)\times S(^{\text{132}}\text{Sn)},
\label{SLSn132b}
\end{equation}%
and
\begin{equation}
S(^{\text{124}}\text{Sn)}=(0.1255\pm 0.0020)+(0.0011\pm 4.05\times
10^{-5})\times L,  \label{SLSn124a}
\end{equation}%
\begin{equation}
L=(-110.1\pm 5.2)+(882.6\pm 32.3)\times S(^{\text{124}}\text{Sn)},
\label{SLSn124b}
\end{equation}%

Similar linear relations between $S$ and $L$ are also expected for
other heavy nuclei. This is not surprising as detailed discussions in Refs. \cite%
{brown00,hor01,typel01,furn02,kara02,diep03} have shown that the
thickness of the neutron skin in heavy nuclei is determined by the
pressure difference between neutrons and protons, which is
proportional to the parameter $L$.

\begin{table}[pt]
\tbl{Linear correlation coefficients $C_{l}$ of $S$
with $L$, $K_{\text{sym}}$ and $E_{\text{sym}}(\protect\rho _{0})$ for $%
^{208}$Pb,\ $^{132}$Sn, $^{124}$Sn, and $^{48}$Ca from 21 sets of
Skyrme interaction parameters.} {\begin{tabular}{@{}ccccc@{}}
\toprule $C_{l}$ $(\%)$ & \quad $^{208}$Pb\quad & $\quad
^{132}$Sn \quad & $^{124}$Sn & $^{48}$Ca \\
\colrule
$S$-$L$\hphantom{00} & \hphantom{0}99.25 & \hphantom{0}98.76 & 98.75 & 93.66 \\
$S$-$K_{\text{sym}}$\hphantom{00} & \hphantom{0}92.26 & \hphantom{0}92.06 & 92.22 & 86.99 \\
$S$-$E_{\text{sym}}$\hphantom{0} & \hphantom{0}87.89 &
\hphantom{0}85.74 & 85.77 & \hphantom{0}81.01\hphantom{0} \\
\botrule
\end{tabular}}
\label{Corr}
\end{table}

To give a quantitative estimate of above discussed correlations,
we define the following linear correlation coefficient $C_{l}$:
\begin{equation}
C_{l}=\sqrt{1-q/t},
\end{equation}%
where%
\begin{equation}
q=\underset{i=1}{\overset{n}{\sum }}[y_{i}-(A+Bx_{i})]^{2},\text{ }t=%
\underset{i=1}{\overset{n}{\sum }}(y_{i}-\overline{y}),~~~\overline{y}=%
\underset{i=1}{\overset{n}{\sum }}y_{i}/n.
\end{equation}%
In the above, $A$ and $B$ are the linear regression coefficients, $(x_{i}$, $%
y_{i})$ are the sample points, and $n$ is the number of sample
points. The linear correlation coefficient $C_{l}$ measures the
degree of linear correlation, and $C_{l}=1$ corresponds to an
ideal linear correlation. Table 1 gives the linear correlation
coefficient $C_{l}$ for the correlation of $S$ with $L$,
$K_{\text{sym}}$ and $E_{\text{sym}}(\rho _{0})$
for $^{208}$Pb, $^{132}$Sn, $^{124}$Sn, and $^{48}$Ca shown in Figs. \ref%
{SPb208} and \ref{SSnCa} for different Skyrme interactions. It is
seen that these correlations become weaker with decreasing nucleus
mass, and a strong linear correlation only exists between the $S$
and $L$ for the heavier nuclei $^{208}$Pb, $^{132}$Sn, and
$^{124}$Sn. Therefore, the neutron skin thickness of these nuclei
can be extracted once the slope parameter $L$ of the nuclear
symmetry energy at saturation density is known.

\section{Predictions on the neutron skin thickness of heavy
nuclei}

The extracted $L$ value from the isospin diffusion data in heavy ion
collisions allows us to determine from Eqs. (\ref{SLPb208a}%
), (\ref{SLSn132a}), and (\ref{SLSn124a}), respectively, a neutron
skin thickness of $0.22\pm 0.04$ fm for $^{208}$Pb, $0.29\pm 0.04$ fm for $^{132}$%
Sn, and $0.22\pm 0.04$ fm for $^{124}$Sn. Experimentally, great
efforts were devoted to measure the thickness of the neutron skin in heavy nuclei \cite%
{sta94,clark03}, and a recent review can be found in Ref.
\cite{kras04}. The data for the neutron skin thickness of
$^{208}$Pb indicate a large uncertainty, i.e., $0.1$-$0.28$ fm.
Our results for the neutron skin thickness of $^{208}$Pb are thus
consistent with present data but give a much stronger constraint.
A large uncertainty is also found experimentally in the neutron skin thickness of $^{124}$Sn, i.e., its value varies from $%
0.1 $ fm to $0.3$ fm depending on the experimental method. The
proposed experiment of parity-violating electron scattering from
$^{208}$Pb at the Jefferson Laboratory is expected to give another
independent and more accurate measurement of its neutron skin
thickness (within $0.05$ fm), thus providing improved constraints
on the density dependence of the nuclear symmetry energy
\cite{hor01b,jeff00}.

Most recently, an accurately calibrated relativistic
parametrization based on the relativistic mean-field theory has
been introduced to study the neutron skin thickness of finite
nuclei \cite{piek05}. This parametrization can describe
simultaneously the ground state properties of finite nuclei and
their monopole and dipole resonances. Using this parametrization,
the authors predicted a neutron skin thickness of $0.21$ fm in
$^{208}$Pb, $0.27$ fm in $^{132}$Sn, and $0.19$ fm in $^{124}$ Sn
\cite{piek05,piek}. These predictions are in surprisingly good
agreement with our results constrained by the isospin diffusion
data from heavy-ion collisions.

\section{Summary}

Both the structure of nuclear surface and the dynamics of isospin
equilibration in heavy ion collisions are affected by the density
dependence of nuclear symmetry energy. From the most recent
analysis of the isospin diffusion data in heavy-ion collisions
using an isospin- and momentum-dependent transport model with
in-medium NN cross sections, a value $L=88\pm 25$ MeV has been
extracted for the slope of the nuclear symmetry energy at
saturation density. This relatively constrained value imposes
strong constraints on the parameters in the Skyrme effective
interactions. Among the 21 sets of commonly used Skyrme
parameters, only SIV, SV, G$_\sigma$, and R$_\sigma$ have $L$
values that are consistent with the extracted one. We have also
studied the correlation between the neutron skin thickness of
finite nuclei and the nuclear symmetry energy within the framework
of the Skyrme Hartree-Fock model. As in previous studies, we have
found a strong linear correlation between the neutron skin
thickness of heavy nuclei and the slope $L$ of the nuclear matter
symmetry energy at saturation density. This correlation provides
stringent constraints on both the density dependence of the
nuclear symmetry energy and the thickness of the neutron skin in
heavy nuclei. The extracted $L$ value from the isospin diffusion
data then leads to predicted
neutron skin thickness of $0.22\pm 0.04$ fm for $^{208}$Pb, $%
0.29\pm 0.04$ fm for $^{132}$Sn, and $0.22\pm 0.04$ fm for
$^{124}$Sn. Our work has thus demonstrated that information
obtained from studying isospin dynamics in heavy ion collisions
can be very useful for understanding the structure of nuclei.

\section*{Acknowledgments}
The work was supported in part by the National Natural Science
Foundation of China under Grant No. 10575071, MOE of China under
project NCET-05-0392, Shanghai Rising-Star Program under Grant No.
06QA14024 (LWC), the US National Science Foundation under Grant
No. PHY-0457265 and the Welch Foundation under Grant No. A-1358
(CMK), as well as the US National Science Foundation under Grant
No. PHY-0354572 and PHY-0456890 and the NASA-Arkansas Space Grants
Consortium Award ASU15154 (BAL).

\end{document}